\documentclass[draft]{agujournal2019}
\usepackage{algorithm}
\usepackage{algpseudocode}
\usepackage{subfigure}
\usepackage[english]{babel}
\usepackage{ragged2e}
\usepackage[none]{hyphenat}
\draftfalse

\journalname{Seismological Research Letters}

\begin{document}
% \nolinenumbers
% \hyphenpenalty=10000
% \exhyphenpenalty=10000
%% ------------------------------------------------------------------------ %%
%  Title
%
% (A title should be specific, informative, and brief. Use
% abbreviations only if they are defined in the abstract. Titles that
% start with general keywords then specific terms are optimized in
% searches)
%
%% ------------------------------------------------------------------------ %%

\title{Real-time processing of distributed acoustic sensing data for earthquake monitoring operations}

%% ------------------------------------------------------------------------ %%
%
%  AUTHORS AND AFFILIATIONS
%
%% ------------------------------------------------------------------------ %%

% Authors are individuals who have significantly contributed to the
% research and preparation of the article. Group authors are allowed, if
% each author in the group is separately identified in an appendix.)

% List authors by first name or initial followed by last name and
% separated by commas. Use \affil{} to number affiliations, and
% \thanks{} for author notes.
% Additional author notes should be indicated with \thanks{} (for
% example, for current addresses).

% Example: \authors{A. B. Author\affil{1}\thanks{Current address, Antartica}, B. C. Author\affil{2,3}, and D. E.
% Author\affil{3,4}\thanks{Also funded by Monsanto.}}

\authors{E. Biondi\affil{1,2}\thanks{1200 E California Blvd, Pasadena, CA 91125}, Gabrielle Tepp\affil{1}, Ellen Yu\affil{1}, Jessie K. Saunders\affil{1}, Victor Yartsev\affil{3},  Michael Black\affil{1}, Michael Watkins\affil{1}, Aparna Bhaskaran\affil{1}, Rayomand Bhadha\affil{1}, Zhongwen Zhan\affil{1}, Allen L. Husker\affil{1}
}

% \affiliation{1}{First Affiliation}
% \affiliation{2}{Second Affiliation}
% \affiliation{3}{Third Affiliation}
% \affiliation{4}{Fourth Affiliation}

\affiliation{1}{California Institute of Technology, Seismological Laboratory; 1200 E. California Blvd., MS 252-21 Pasadena, California 91125-2100}

\affiliation{2}{Stanford University, Geophysics Department; 397 Panama Mall Mitchell Building, Stanford, California, 94305}

\affiliation{3}{Luna Innovations, Chino, California, United States}

%% Corresponding Author:
% Corresponding author mailing address and e-mail address:

% (include name and email addresses of the corresponding author.  More
% than one corresponding author is allowed in this LaTeX file and for
% publication; but only one corresponding author is allowed in our
% editorial system.)

% Example: \correspondingauthor{First and Last Name}{email@address.edu}

\correspondingauthor{Ettore Biondi}{ettore88@stanford.edu}

%% Keypoints, final entry on title page.

%  List up to three key points (at least one is required)
%  Key Points summarize the main points and conclusions of the article
%  Each must be 100 characters or less with no special characters or punctuation and must be complete sentences

% Example:
% \begin{keypoints}
% \item	List up to three key points (at least one is required)
% \item	Key Points summarize the main points and conclusions of the article
% \item	Each must be 100 characters or less with no special characters or punctuation and must be complete sentences
% \end{keypoints}

% \begin{keypoints}
% \item enter point 1 here
% \item enter point 2 here
% \item enter point 3 here
% \end{keypoints}

%% ------------------------------------------------------------------------ %%
%
%  ABSTRACT and PLAIN LANGUAGE SUMMARY
%
% A good Abstract will begin with a short description of the problem
% being addressed, briefly describe the new data or analyses, then
% briefly states the main conclusion(s) and how they are supported and
% uncertainties.

% The Plain Language Summary should be written for a broad audience,
% including journalists and the science-interested public, that will not have 
% a background in your field.
%
% A Plain Language Summary is required in GRL, JGR: Planets, JGR: Biogeosciences,
% JGR: Oceans, G-Cubed, Reviews of Geophysics, and JAMES.
% see http://sharingscience.agu.org/creating-plain-language-summary/)
%
%% ------------------------------------------------------------------------ %%

%% \begin{abstract} starts the second page

\begin{abstract}
\justifying
We introduce a modular software framework designed to integrate distributed acoustic sensing (DAS) data into operational earthquake monitoring systems. Building on the infrastructure of the Advanced National Seismic System (ANSS) and the Southern California Seismic Network (SCSN), which employs the ANSS Quake Monitoring Software (AQMS), our solution supports real-time DAS waveform streaming and machine-learning-based traveltime picking to leverage the dense spatial sampling of DAS arrays. To enable seamless compatibility with the AQMS, our approach uses standardized seismic data formats to incorporate predetermined DAS channels. We demonstrate the integration of data from a 100-km-long DAS array deployed in Ridgecrest, California, and provide a detailed description of the software components and deployment strategy. This work represents a step toward incorporating DAS into routine seismic monitoring and opens new possibilities for real-time hazard assessment using fiber-optic networks.
\end{abstract}

\section*{Introduction}
Since the first recordings of large-magnitude earthquakes using seismometers in the late 1800s~\cite{von1889earthquake,national2003living} to the establishment of the National Strong-Motion Program in the 1930s~\cite{haddadi2008center}, earthquake monitoring operations have undergone significant advancements~\cite{allen2009status,allen2019earthquake}. These efforts are typically coordinated by organized seismic networks~\cite{kanamori2005real}, which are responsible for maintaining and operating groups of seismic stations within defined regions. In the United States, the vision of a unified national seismic system developed between 1980 and 2000, culminating in the creation of the Advanced National Seismic System (ANSS)~\cite{benz2001advanced}. ANSS integrates national, regional, and local seismic monitoring initiatives, delivering comprehensive data to support earthquake research and enhance public safety. The Southern California Seismic Network (SCSN) has been operating seismic stations and monitoring Southern California seismicity since the 1930s~\cite{https://doi.org/10.7914/sn/ci}. 

The advent of fiber-optic instruments for seismological applications has recently opened new avenues for research~\cite{zhan2020distributed,lindsey2021fiber,cheng2024photonic}. In particular, distributed acoustic sensing (DAS) systems have driven major progress across multiple scientific domains~\cite{jousset2018dynamic,zhu2019characterizing,landro2022sensing,yang2022sub,ajo2022imperial,li2023break,biondi2023upper,manos2024discharge,liu2025urban,li2025daily,fichtner2025hidden}. By transforming long stretches of telecommunication fiber into dense seismic arrays, DAS provides observational capabilities that significantly complement and expand those of traditional seismic stations. Moreover, modern DAS instruments are able to perform sensing on multiple fiber strands and without disrupting telecommunication operations~\cite{mazur2024global,shi2025multiplexed}, opening new opportunities for the seamless integration of sensing operations within the data transmission by fiber networks and allowing broad, network-scale deployments~\cite{farghal2022potential,yin2023real,gou2025leveraging,mcguire2025fiber}.

Despite its promise, DAS introduces new challenges due to the vast number of channels recorded—often resulting in tens of gigabytes of data per day. This data volume presents significant hurdles in terms of real-time management and processing. Moreover, traditional seismic network operations are typically designed around independent, sparsely distributed stations and are not optimized for the continuous, high-density data streams generated by DAS. While recent efforts have produced software tools capable of handling large DAS datasets~\cite{chambers2024dascore,ni2024object}, there remains a critical need for modular, scalable code infrastructures that can include DAS into existing monitoring operations.

For decades, the SCSN has relied on conventional seismic stations for earthquake monitoring~\cite{hauksson2001southern,hutton2010earthquake}. The incorporation of DAS data, particularly from submarine cables, has the potential to significantly enhance our understanding of Southern California's seismicity. We present a new workflow and its corresponding Python-based software implementation that enables the seamless integration of selected DAS channels into the existing ANSS Quake Monitoring Software (AQMS), which is used by some regional networks for operational monitoring. Our approach leverages standardized seismic data formats to support real-time processing of DAS data streams, as well as the ingestion of traveltime picks derived from a machine-learning algorithm specifically designed to exploit the high spatial density of DAS arrays~\cite{zhu2023seismic}. The modular, object-oriented architecture of our code provides a flexible foundation that can be adapted for other DAS real-time processing paradigms within seismic network operations. 

We begin by outlining the conceptual framework for incorporating DAS data into the ongoing earthquake monitoring activities of the SCSN. We then demonstrate the ingestion of data from an active DAS array currently deployed in Ridgecrest, California. Finally, we detail the main components of our software infrastructure and provide guidance on its operation. 

\section*{Real-time DAS data streaming and processing}
\begin{figure}[ht]
\centering
\includegraphics[width=0.9\linewidth]{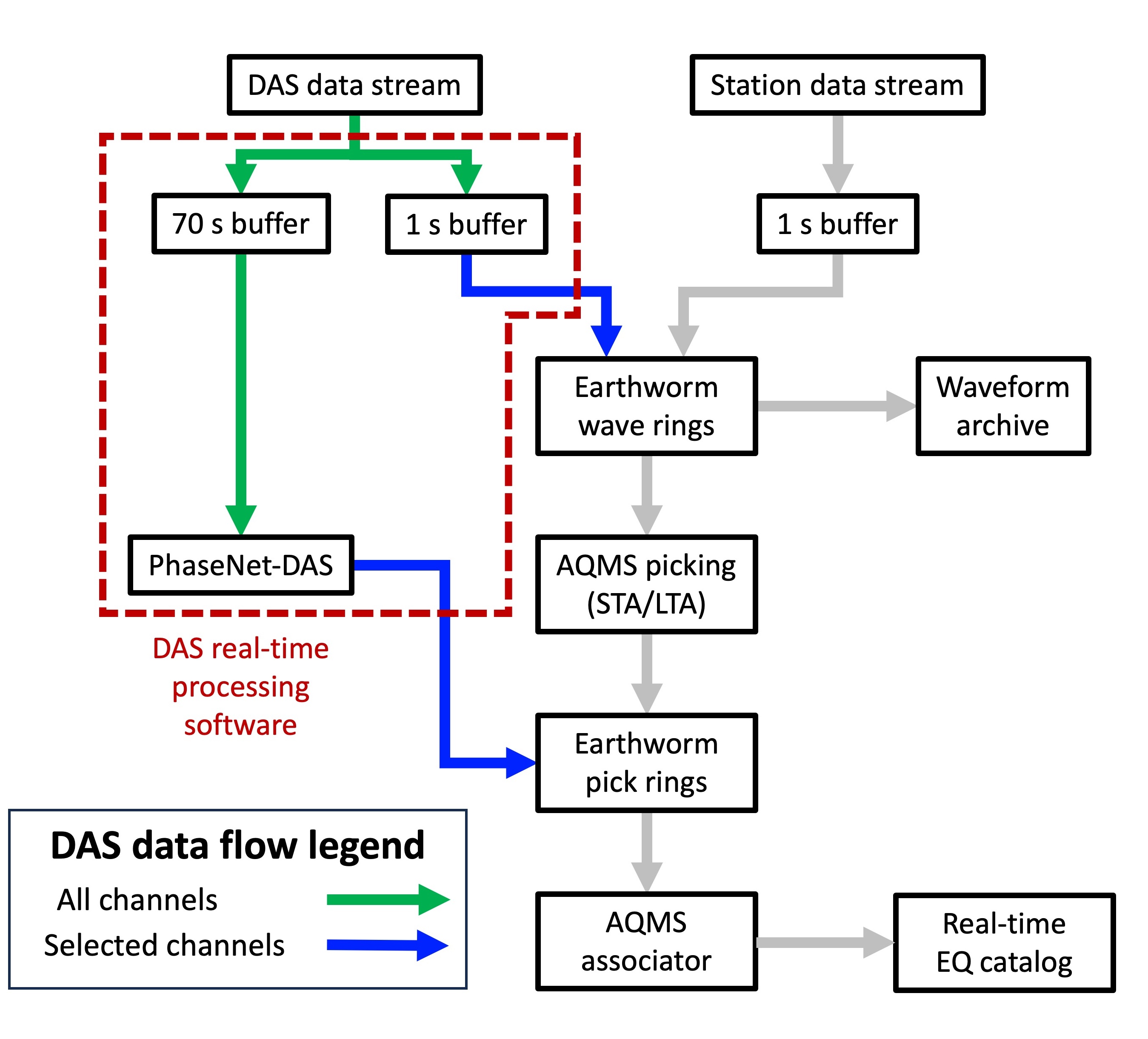}
\caption{Schematic describing the currently operating data and processing flows for DAS and seismic stations for earthquake monitoring operations at the SCSN. AQMS = ANSS Quake Monitoring System; DAS = Distributed Acoustic Sensing; EQ = earthquake}
\label{fig:figure1}
\end{figure}

The SCSN monitors regional seismicity using more than 500 sensors distributed across Southern California. Each sensor transmits data to Caltech in 1-second packets, with the number of time samples per packet depending on the sensor's sampling rate (typically 40 or 100 Hz). A schematic overview of the data flow is shown in Figure~\ref{fig:figure1}, illustrating how these data packets are integrated into the AQMS~\cite{renate2020open}. Within AQMS, the Earthworm framework handles the ingestion of packets into ring buffers to enable real-time processing~\cite{olivieri2012almost}. The waveform data are simultaneously directed to both the SCSN archive~\cite{center2013southern} and the real-time earthquake monitoring pipeline, which includes phase picking and event association to generate earthquake catalogs. Currently, AQMS uses the short-term average/long-term average (STA/LTA) method for event detection~\cite{trnkoczy2009understanding}, but efforts are made to extend AQMS to use machine-learning algorithms to improve picking sensitivity and accuracy~\cite{retailleau2022wrapper, tepp2025ml}.

To seamlessly integrate data from Distributed Acoustic Sensing (DAS) instruments into the existing AQMS infrastructure, we adopt a strategy analogous to that used for conventional seismic stations (Fig.~\ref{fig:figure1}). DAS data are streamed from the interrogator using a WebSocket protocol and ingested by our custom software component. This software assigns predetermined channel identifiers and inserts the data into 1-second Earthworm buffer rings, allowing them to be treated as standard seismic traces within the AQMS workflow. Given that DAS arrays typically consist of thousands of channels spanning tens of kilometers, incorporating all channels into real-time monitoring would create a significant imbalance relative to traditional seismic stations. Therefore, only a subset of well-coupled DAS channels should be selected for operational monitoring. We follow a strategy where channels are spaced at least 5 km apart to minimize redundancy and preserve computational efficiency. This value was chosen to have channel density comparable with SCSN regions with dense station coverage. These selected channels can then be processed by AQMS to contribute to earthquake detection across the monitored region.

While this strategy offers a straightforward way to ingest a subset of information from DAS arrays, it does not fully exploit the rich data content DAS provides. To leverage the full potential of DAS for earthquake monitoring, we employ PhaseNet-DAS, a recently developed deep-neural-network-based picking algorithm tailored for DAS data~\cite{zhu2023seismic}. We apply this machine learning model in real-time, enabling the processing of DAS data as they are streamed into a rolling 70-second ring buffer. Once the buffer is filled, PhaseNet-DAS is applied to extract phase picks from all channels across the processed DAS array as quickly as possible to achieve the highest processing throughput, which in our tests is approximately 1 second per picking task.

The resulting traveltime picks from selected DAS channels are then inserted into an Earthworm pick ring, allowing seamless injection into the AQMS phase associator. In this configuration, the DAS waveform data are archived directly from the 1-second buffer but are excluded from the AQMS internal picking process, which is instead handled by PhaseNet-DAS externally. The PhaseNet-DAS code executes continuously, initiating a new inference cycle as soon as the previous one completes; thus, the processing speed is governed by the number of DAS channels and the length of the 70-second data segment. To avoid redundant detections, the system filters out picks that are within 1 second of any previously transmitted pick, ensuring that only newly identified arrivals are streamed.

\subsection*{Incorporating the Ridgecrest DAS array into SCSN monitoring}
\begin{figure}[ht]
\centering
\includegraphics[width=0.9\linewidth]{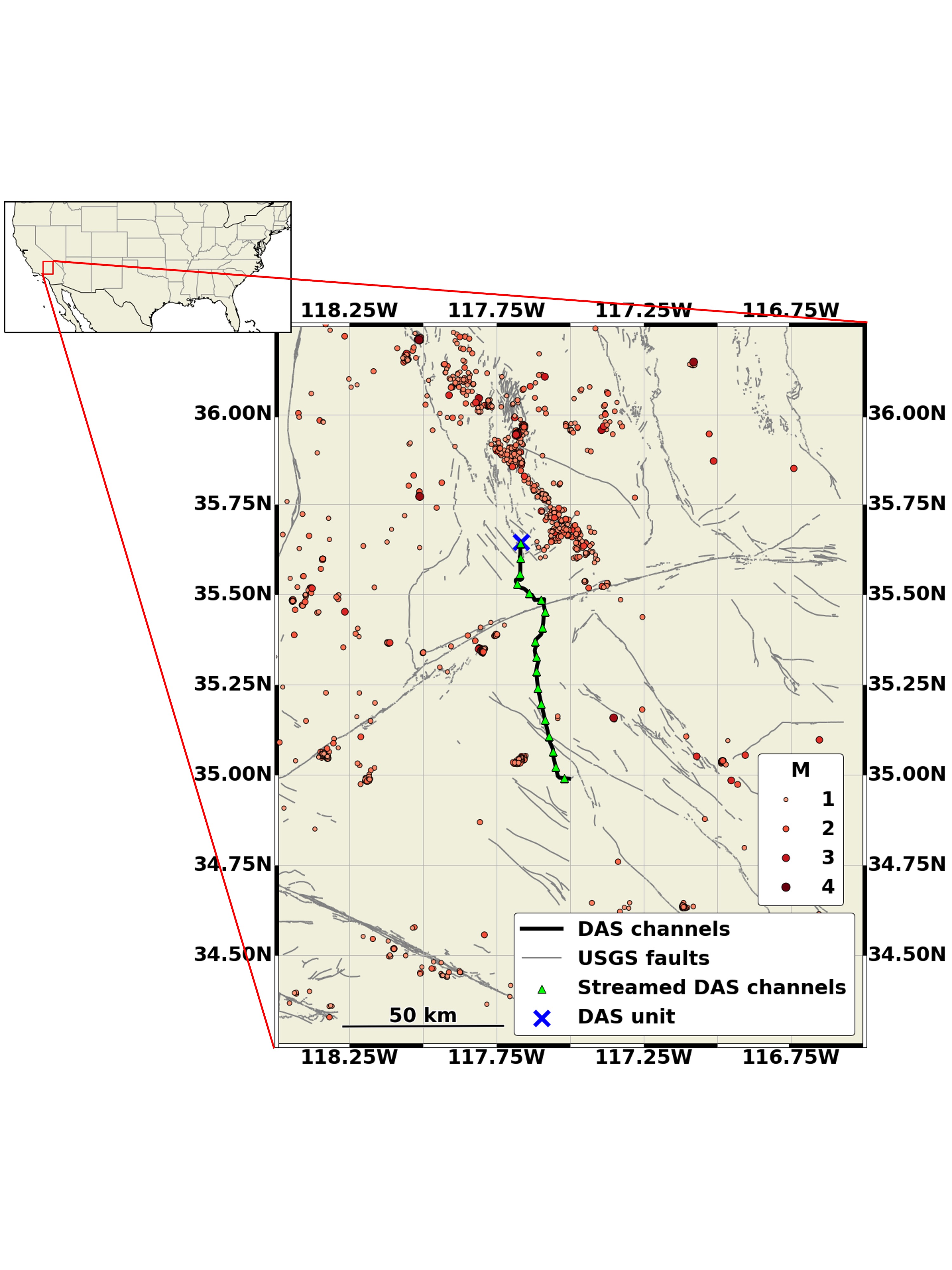}
\caption{Location of the Ridgecrest DAS array (black line) with the considered DAS channels currently being tested for earthquake monitoring operations (green triangles). The instrument deployment position is depicted by the blue cross. The gray lines represent the local faults within the United States Geological Survey (USGS) catalog~\cite{frankel2000usgs}, and the red dots depict the local seismicity that occurred between August 2024 (the start of continuous real-time DAS streaming) and April 2025 from the SCSN catalog.}
\label{fig:figure2}
\end{figure}

To demonstrate the effectiveness of our DAS ingestion workflow, we test the proposed approach by streaming data from a DAS array located in Ridgecrest, CA (Fig.~\ref{fig:figure2}). This array has enabled detailed imaging of structures within the Garlock Fault and the Moho interface in Southern California~\cite{atterholt2024imaging, atterholt2024fine}. Additionally, a shorter DAS array in the same area increased the number of detected aftershocks of the 2019 M7.1 Ridgecrest earthquake by two orders of magnitude, using a template matching approach compared to the conventional catalog~\cite{ross2019hierarchical,li2021rapid}. These results suggest that DAS arrays can enhance earthquake monitoring operations and knowledge of subsurface structures by providing supplemental seismic data from this region.

The Ridgecrest DAS array comprises 10,000 channels spanning an effective length of 80 km, with 10-meter spatial sampling after excluding fiber loops and poorly coupled channels. Channel locations were obtained using the vehicle-based geolocation methodology described by \citeA{biondi2023geolocalization}. The system’s ping rate is set to 1 kHz, while data are stored and streamed at a downsampled rate of 100 Hz, similar to conventional seismic stations. A gauge length of 100 meters is used to improve the signal-to-noise ratio (SNR) of distant channels affected by fading instrumental noise.

\subsubsection*{Data streaming and selected channel metadata}
From the 10,000 available channels, we stream 5,000 via a WebSocket connection to a remote processing server located on the Caltech campus. The number of streamed channels is limited to half the total due to bandwidth constraints. For real-time testing relevant to earthquake monitoring operations, we select 18 channels evenly spaced along the array, with an inter-channel spacing of approximately 5~km. The instrument transmits individual data packets at 100~Hz, with each packet containing optical phase measurements for all 5,000 streamed channels. These packets are received by our custom software running on a dedicated server at Caltech. The system has been streaming data continuously since August 2024, with an average latency of 0.6 seconds per packet and a telemetry reliability exceeding 99.989\%. Moreover, given the large bandwidth provided by telecommunication fibers, we can also download the complete 10,000 channel files with higher latency (i.e., 2 minutes per 5-minute-long file). Since the data are stored within the waveform database, the network analysts can access and manually pick the seismic DAS traces using the AQMS remote client Jiggle~\cite{renate2020open} (Fig. S1).

To keep track of seismic trace metadata and instrument changes, the Standard for the Exchange of Earthquake Data (SEED) format was introduced~\cite{ringler2015quick}. Similar efforts are currently evolving around the definition of metadata standards for DAS traces~\cite{lai2024toward}; however, the community has not yet reached a general consensus nor achieved widespread adoption of a unified standard. Since our approach injects only a minimal portion of the DAS traces into the AQMS system, we adopt the SEED definition to represent the selected DAS channels. For instance, a given channel from the Ridgecrest DAS array is labeled \texttt{CI.DRS02..HS1}, with the following definitions:

\begin{itemize}
  \item \textbf{CI} = network code (CI stands for SCSN).
  \item \textbf{DRS02} = station code; DAS Ridgecrest South (DRS) channel 02 (arbitrary channel counter allowing up to 100 channels).
  \item \textbf{[currently blank]} = location code; we may assign non-blank location codes for future experimentation, such as stacking nearby DAS channels to improve SNR and/or including adjacent channels.
  \item \textbf{HS1} = channel code; where, \textbf{H} = high rate (100 Hz), \textbf{S} = strain, and \textbf{1} = arbitrary strain component.
\end{itemize}

Using the SEED format to define specific DAS channels allows us to monitor data ingestion via existing seismic network systems~\cite{hauksson2001southern}. Regarding DAS instrument response, current understanding suggests that DAS instruments exhibit a flat response within the seismic frequency band~\cite{paitz2021empirical, lindsey2020broadband}. Based on this assumption, the phase changes recorded by a DAS unit can be directly converted into strain rate (1/s). In our system, we convert raw phase values into microstrain rate ($\mu$m/m/s) and store them as waveforms in the network’s archive. This assumption may be revised, as emerging research indicates a minor frequency-dependent response in such systems~\cite{zhai2025comprehensive, chien2025calibrating}, potentially related to fiber-optic cable installation.

\subsubsection*{Real-time processing and earthquake sequence example}
For the Ridgecrest array, we collect data packets into the previously described 70-second rolling window, which is processed in real time by PhaseNet-DAS to obtain earthquake traveltime picks. By applying this tool to all 5,000 streaming channels, we fully leverage the potential of DAS for earthquake detection, as opposed to relying on traditional trace-by-trace algorithms~\cite{zhu2023seismic}. The picks obtained by PhaseNet-DAS from the 18 channels are then placed within an Earthworm pick ring to be ingested in real time by the AQMS workflow. To do so, we employ the Python interface for Earthworm developed by~\citeA{hernandez2018new}, which allows sending such picks as formatted string messages. At the time of writing, 8 channels out of the selected 18 channels are currently being used into the SCSN monitoring operations.

Figures~\ref{fig:figure3}(a) and (b) show examples of PhaseNet-DAS traveltime picks from local magnitude 3.4 and 1.0 earthquakes recorded on the Ridgecrest array with distances of 95 and 52 km from the array center, respectively. In both cases, red and blue dots represent P- and S-wave picks. For the $M$3.4 event, the P- and S-wave picks clearly follow the wavefront arrival. Only the P-wave picks at the furthest channels are affected by instrument noise. Additionally, a secondary phase arriving after the main S-wave and recorded in the second half of the array is also labeled as an S-wave. In contrast, for the smaller $M$1.0 event (Fig.~\ref{fig:figure3}b), PhaseNet-DAS detects only part of the wavefronts, and some P-wave arrivals between channels 2400 and 3000 are misclassified. These mislabeled picks could be removed using a phase association algorithm during a pick preprocessing phase~\cite{zhu2022earthquake}, which is currently not part of our software, but it can be included after the picking process has finished. The performance of PhaseNet-DAS on a rolling window is shown in Supplementary Movie S1.

Figure~\ref{fig:figure3new} depicts the measured signal-to-noise ratio (SNR) based on the P- and S-wave picks obtained by PhaseNet-DAS and magnitude distribution for nearly 250 events recorded within 120~km of the DAS array center that occurred between May 27 and July 31, 2023. The SNR is calculated using energy windows of 2 seconds before and 0.5 seconds after the traveltime picks for noise and signal, respectively. The SNR increases with earthquake magnitude, with a more pronounced improvement for P-wave picks. S-wave picks exhibit a less pronounced increase due to the influence of near-surface scattered P-wave coda waves. 

To visually assess the performance of PhaseNet-DAS for real-time picking, we analyze the traveltime picks obtained by our workflow for the Lamont M5.2 earthquake sequence, which occurred approximately 150 km east of the Ridgecrest DAS array (Fig.\ref{fig:figure4}a), near the White Wolf fault~\cite{stein1981seismic}. The mainshock of this sequence occurred on August 8, 2024, and generated more than 400 aftershocks detected by our real-time DAS earthquake picking system. Figure~\ref{fig:figure4}b shows the DAS strain rate recorded by the 5000 streamed channels, along with the P- and S-wave picks obtained by our real-time PhaseNet-DAS workflow. These picks are in good agreement with the detections from nearby SCSN stations. The Supplementary Movie S2 shows the performance of PhaseNet-DAS for this earthquake sequence. In contrast, a comparison with a conventional STA/LTA threshold picking algorithm (Fig.~\ref{fig:figure4}c) reveals that STA/LTA misses a significant number of aftershocks and introduces multiple mispicks, primarily from moving vehicles. Although lowering the STA/LTA threshold could improve the detection of weaker aftershocks, it would also substantially increase the number of false picks resulting from traffic and instrument noise (Fig. S2).

\subsubsection*{Discussion on the inclusion of DAS amplitude information}
The Earthworm pick ring paradigm employed here is capable of ingesting converted strain-rate amplitudes once an appropriate calibration function is developed. Recent efforts have focused on understanding the relationship between ground motion and the strain rates recorded by DAS instruments deployed on telecommunication cables~\cite{lindsey2020broadband,lior2024accurate,fairweather2024characterisation,sawi2024assessing,hudson2025unlocking,zhai2025comprehensive,chien2025calibrating}. In addition, magnitude scaling relationships specific to DAS data have been proposed and validated~\cite{yin2023earthquake,nayak2024seismic,strumia2024sensing}, further demonstrating the quantitative reliability of strain-rate measurements on dark fiber strands. Nevertheless, uncertainties remain regarding the coupling and dynamic range of DAS systems, particularly for large-magnitude events~\cite{van2024analysis}. Therefore, amplitude information is currently excluded from our system, but the workflow is designed to flexibly incorporate such data once the relationships between ground motion and strain are fully validated and tested. 

\section*{Code description}
In this section, we provide an overview of the main components of the DAS processing software and demonstrate how to correctly run the code. The package has been implemented using Python and the installation is managed through a \textit{conda} environment to ensure easy distribution and cross-platform compatibility~\cite{maji2020demystifying}. All operational modes are controlled by a central script, \textit{StreamProcessor.py}, which connects the processing server to the streaming data socket and collects packets into rolling buffers for either PhaseNet-DAS-based picking or Earthworm waveform processing. The code is capable of handling streamed strain or strain rate data, packaged either as individual time samples or as multiple time samples across all channels. The main structure of the driving script is shown in algorithm~\ref{alg:das-stream}, which can help a user read the implemented software.  

\begin{algorithm}
\caption{Real-time DAS Stream Processing}\label{alg:das-stream}
\begin{algorithmic}[1]
\State \textbf{Input:} Stream parameters (\texttt{host}, \texttt{port}, etc.), processing settings (\texttt{fs}, \texttt{workInterval}, etc.)
\State \textbf{Initialize} StreamReader, ringbuffer, and conversion factors
\State \textbf{Connect to stream source}
\While{connection is active}
    \State \textbf{Try} to receive next packet
    \If{packet is empty}
        \State \textbf{Break}
    \EndIf
    \State Extract timestamp and payload
    \If{time gap with last sample is inconsistent}
        \State Log skipped samples
        \State \textbf{Optionally:} write buffer to disk
        \State \textbf{Break}
    \EndIf
    \State Append data and timestamp to ring buffer
    \If{ringbuffer has reached full rotation}
        \State Optionally send data to Earthworm wavering if defined
    \EndIf

    \If{buffer has reached \texttt{workInterval} duration}
        \If{picking enabled}
            \If{previous pick task is complete}
                \State Retrieve new picks
                \If{Earthworm pickring is defined}
                    \State Stream new picks to pickring
                \EndIf
                \State Launch asynchronous picking task
            \EndIf
        \EndIf
    \EndIf
\EndWhile
\State \textbf{Catch} and log exceptions
\end{algorithmic}
\end{algorithm}

\subsection*{Stream Reader template}
To support different paradigms and packet formats across various DAS units, the code adopts an object-oriented design for collecting and unpacking streamed DAS data. This functionality is encapsulated in a template class named \textit{StreamReader}. The core method of this class, \textit{getNextPacket}, receives the next data packet from the socket and unpacks its contents, which can then be accessed through additional class methods. For example, \textit{getPacketTimestamp} retrieves the timestamps of the streamed samples, while \textit{getPayloadRad} returns the actual streamed data values. These two methods represent the primary interfaces used in the main processing scripts.

Additional templated methods are provided to access relevant acquisition parameters, such as sampling rate, decimation frequency, and gauge length. In the current implementation, we demonstrate how data from the Alcatel Submarine Network OptoDAS unit are collected and unpacked using this framework.

The data collected by the \textit{StreamReader} object are stored in another class called \textit{RingBuffer}. This class manages the concatenation of streamed packets into rolling buffers of a predetermined size, suitable for ingestion into Earthworm rings or for real-time picking by PhaseNet-DAS. The PhaseNet-DAS process runs asynchronously in a separate thread, while the main thread continues to handle data collection. The modular architecture of the software also facilitates the integration of additional real-time processing modules tailored to other applications, such as volcano monitoring~\cite{sparks2012monitoring,doi:10.1126/science.adu0225}.

\subsection*{Channel selection}
As previously described, only a small subset of channels is ingested by the network operations. For the Ridgecrest array, we test the inclusion of 18 selected channels, evenly spaced at approximately 5~km intervals. These channels are specified using a standardized eXtensible Markup Language (XML) station file, which is read by the main processing script. This XML-formatted file can be used to read the channel information using the well-established ObsPy Python package~\cite{beyreuther2010obspy}.

The XML file is generated by a helper script named \textit{CreateMetaData.py}, which takes a comma-separated values (CSV) file as input. Each row in the CSV includes the channel index, a flag indicating whether the channel is well-coupled, and the corresponding latitude, longitude, and elevation. Additional arguments to the script allow the user to define the set of streamed channels available from the instrument, select the subset to be ingested by the Earthworm workflow, and assign standardized SEED metadata parameters to each selected channel.

For instance, for the Ridgecrest array, we create the XML file with the following Bash command:

\begin{verbatim}
CreateMetaData.py \ 
    Dat/Ridgecrest-DAS-100km.csv \ #CSV file containing DAS channel positions
    -dsmp 250 \                    #Downsampling factor for selecting channels 
    -dsmpstream 2 \                #Downsampling factor when streaming 
    --GaugeL 102.09524 \           #Gauge length of the DAS system in meters
    --ChSamp 10.209524 \           #Channel spacing in meters
    --fs 100.0 \                   #Sampling frequency in Hz
    --manufacturer OptaSense \     #Manufacturer of the DAS interrogator unit
    --model Plexus \               #Model of the DAS interrogator unit
    --source SCSN \                #Source organization name 
    --network CI \                 #SEED standard network code
    --array_name RS \              #Array or station name prefix
    --description "Ridgecrest DAS channel" \#Description for each channel
    --output Meta/DAS_RidgecrestSouth100km.xml
\end{verbatim}

In this case, the original 10,000 channels are downsampled by a factor of 2 before streaming them to the processing server at Caltech. Out of the remaining 5,000, we select every 250th channel, which corresponds to a spacing of approximately 5~km between selected channels.

\subsection*{Usage examples}
To show how our workflow can be used, we report the two modalities of how we ingest the waveform and traveltime picks from the Ridgecrest array. To stream the data into the AQMS data workflow, we employ the following command line syntax:

\begin{verbatim}
StreamProcessor.py \
    --host ${IP} \                   % IP address of the data source
    --port ${port} \                 % Port number for the data stream
    -strTp OptaSense \               % Stream type, e.g., OptaSense packet format
    --ringbuffer 1.0 \               % Size of the ring buffer in seconds
    -strnRt 1 \                      % Compute realt-time strain rate from strain
    --xmlmeta Meta/DAS_RidgecrestSouth100km.xml \ % Path to XML metadata file
    -wring 1005 8 141 30 0           % wavering configuration parameters
\end{verbatim}

The data streaming socket is defined by the remote server IP address and a preselected port, which allows the remote server to collect data in real time. A simple string is used to determine the packet format; in this case, it is \texttt{OptaSense}. The data from the selected channels—specified via the XML metadata file—are collected into a 1-second ring buffer and finally injected into an Earthworm wavering, which can be ingested by the operational AQMS earthquake monitoring workflow at SCSN. Since the same socket can be used to also enable real-time traveltime picking using PhaseNet-DAS, we call the same script with slightly different parameters as follows:

\begin{verbatim}
StreamProcessor.py \
    --host ${IP} \                   % IP address of the data source
    --port ${port} \                 % Port number for the data stream
    -strTp OptaSense \               % Stream type, e.g., OptaSense packet format
    -wrkint 1.0 \                    % Picking interval in seconds
    --ringbuffer 70.0 \              % Size of the ring buffer in seconds
    -strnRt 1 \                      % Compute realt-time strain rate from strain
    --xmlmeta Meta/DAS_RidgecrestSouth100km.xml \ % Path to XML metadata file
    -pick 1 \                        % Enable PhaseNet-DAS picking
    -pring 1010 8 141 30 0           % pickring configuration parameters
\end{verbatim}

In this case, the streaming data are collected into a 70-second window, and we enable the picking processing using a boolean flag. To ingest the traveltime picks into AQMS, we provide the necessary \texttt{pyEarthworm} parameters for a predetermined pick ring. The work interval is set to 1 second, which defines how frequently the system checks whether the previous picking task has completed; if so, it asynchronously applies PhaseNet-DAS to the new data window.
In both data ingestion modes, the OptaSense unit streams strain measurements. However, since PhaseNet-DAS is trained using strain rate, we enable the real-time evaluation of strain rate values from the incoming strain data.

Finally, following the same strategy adopted for other real-time operations within SCSN, we run these workflows on two geographically separate servers to ensure system redundancy. By using a scripting-based methodology, this framework can be deployed on the network's premises, in the cloud, or at the edge, enabling flexible integration of DAS real-time data processing into various operational environments.

\section*{Conclusions}
We present a software infrastructure that enables the seamless ingestion of real-time DAS data into seismic networks employing the AQMS workflow. Our solution supports streaming of data packets into existing Earthworm wavering systems, as well as ingestion of traveltime picks obtained using the machine-learning tool PhaseNet-DAS. The latter leverages the large scale and high spatial channel density of DAS arrays deployed along telecommunication cables. We describe the general framework of the system and report on its testing using data from a 100-km DAS array located in Ridgecrest, California. Finally, we outline the main components of the proposed Python-based code and provide an example illustrating its use for real-time data streaming and seismic phase picking. 

Our modular infrastructure marks a step toward the integration of DAS into routine earthquake monitoring operations. Designed with adaptability in mind, the system can be extended to support other DAS deployments, sensor configurations, or machine learning models. By bridging the gap between advanced fiber sensing technologies and operational seismic workflows, this framework opens the door to new applications in real-time hazard monitoring, where DAS and traditional seismic stations can be effectively combined. While our approach advances the incorporation of DAS into existing seismic networks, it also underscores the need to explore alternative paradigms that can fully harness the unique capabilities of DAS. Such exploration will be essential as the vision of large-scale, fiber-based seismic sensing continues to take shape.

\bibliography{main}

\section*{Acknowledgments}
The authors also thank the California Broadband Cooperative for providing access to the Digital 395 telecommunication fibers and their constant support in permitting continuous monitoring operations. This work was funded by the United States Geological Survey National Earthquake Hazards Reduction Program award number G24AP00060 and the California Governor's Office of Emergency Services.

\section*{Author contributions statement}

Conceptualization: EB, EY, JS, RB, GT, AH
Methodology and software development: EB, VY, JS, GT, EY, MB, MW, AB
Data collection: EB, ZZ
Visualization: EB
Supervision: AH
Writing—original draft: EB
% Writing—review \& editing: 

% E. Biondi\affil{1}\thanks{1200 E California Blvd, Pasadena, CA 91125}, Gabrielle Tepp\affil{1}, Ellen Yu\affil{1}, Jessie K. Saunders\affil{1}, Victor Yartsev\affil{2},  Micheal Black\affil{1}, Micheal Watkins\affil{1}, Aparna Bhaskaran\affil{1}, Rayomand Bhadha\affil{1}, Zhongwen Zhan\affil{1}, Allen L. Husker\affil{1}

\subsection*{Contact information}
\begin{itemize}
    \item {\bf{Ettore Biondi}}: California Institute of Technology, Seismological Laboratory; 1200 E. California Blvd., MS 252-21 Pasadena, California 91125-2100
    \item {\bf{Gabrielle Tepp}}: California Institute of Technology, Seismological Laboratory; 1200 E. California Blvd., MS 252-21 Pasadena, California 91125-2100
    \item {\bf{Ellen Yu}}: California Institute of Technology, Seismological Laboratory; 1200 E. California Blvd., MS 252-21 Pasadena, California 91125-2100
    \item {\bf{Jessie K. Saunders}}: California Institute of Technology, Seismological Laboratory; 1200 E. California Blvd., MS 252-21 Pasadena, California 91125-2100
    \item {\bf{Victor Yartsev}}: Luna Innovations; 14351 Pipeline Ave \#5642, Chino, California 91710
    \item {\bf{Michael Black}}: California Institute of Technology, Seismological Laboratory; 1200 E. California Blvd., MS 252-21 Pasadena, California 91125-2100
    \item {\bf{Michael Watkins}}: California Institute of Technology, Seismological Laboratory; 1200 E. California Blvd., MS 252-21 Pasadena, California 91125-2100
    \item {\bf{Aparna Bhaskaran}}: California Institute of Technology, Seismological Laboratory; 1200 E. California Blvd., MS 252-21 Pasadena, California 91125-2100
    \item {\bf{Rayomand Bhadha}}: California Institute of Technology, Seismological Laboratory; 1200 E. California Blvd., MS 252-21 Pasadena, California 91125-2100
    \item {\bf{Zhongwen Zhan}}: California Institute of Technology, Seismological Laboratory; 1200 E. California Blvd., MS 252-21 Pasadena, California 91125-2100
    \item {\bf{Allen L. Husker}}: California Institute of Technology, Seismological Laboratory; 1200 E. California Blvd., MS 252-21 Pasadena, California 91125-2100
\end{itemize}

\section*{Data and Resources}
The software for real-time DAS processing can be found at this GitHub repository:

https://github.com/biondiettore/DAS-realtime.git

\section*{Additional information}

\subsection*{Competing interests}
The authors declare no competing interests.

\begin{figure}[ht]
\centering
\includegraphics[width=1.0\linewidth]{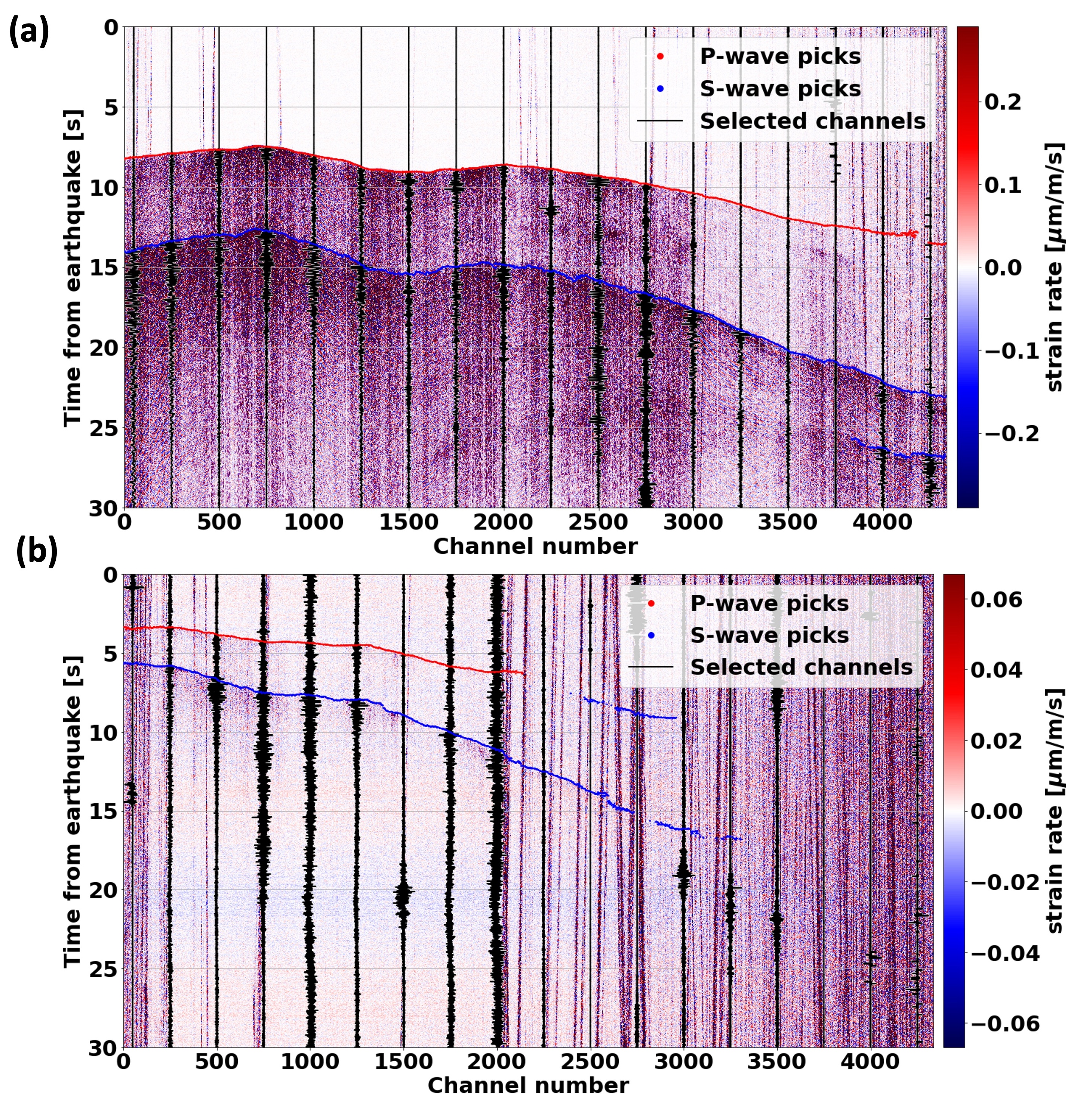}
\caption{PhaseNet-DAS picking examples for a local M3.4 (event ID: 72110903, distance from array 52 km) (a) and an M1.0 (event ID: 72130223, distance from array 40 km)(b). P- and S-wave picks are depicted by the red and blue dots, respectively. The black lines show the traces of the selected DAS channels whose amplitudes are normalized by their maximum for visualization purposes.}
\label{fig:figure3}
\end{figure}

\begin{figure}[ht]
\centering
\includegraphics[width=1.0\linewidth]{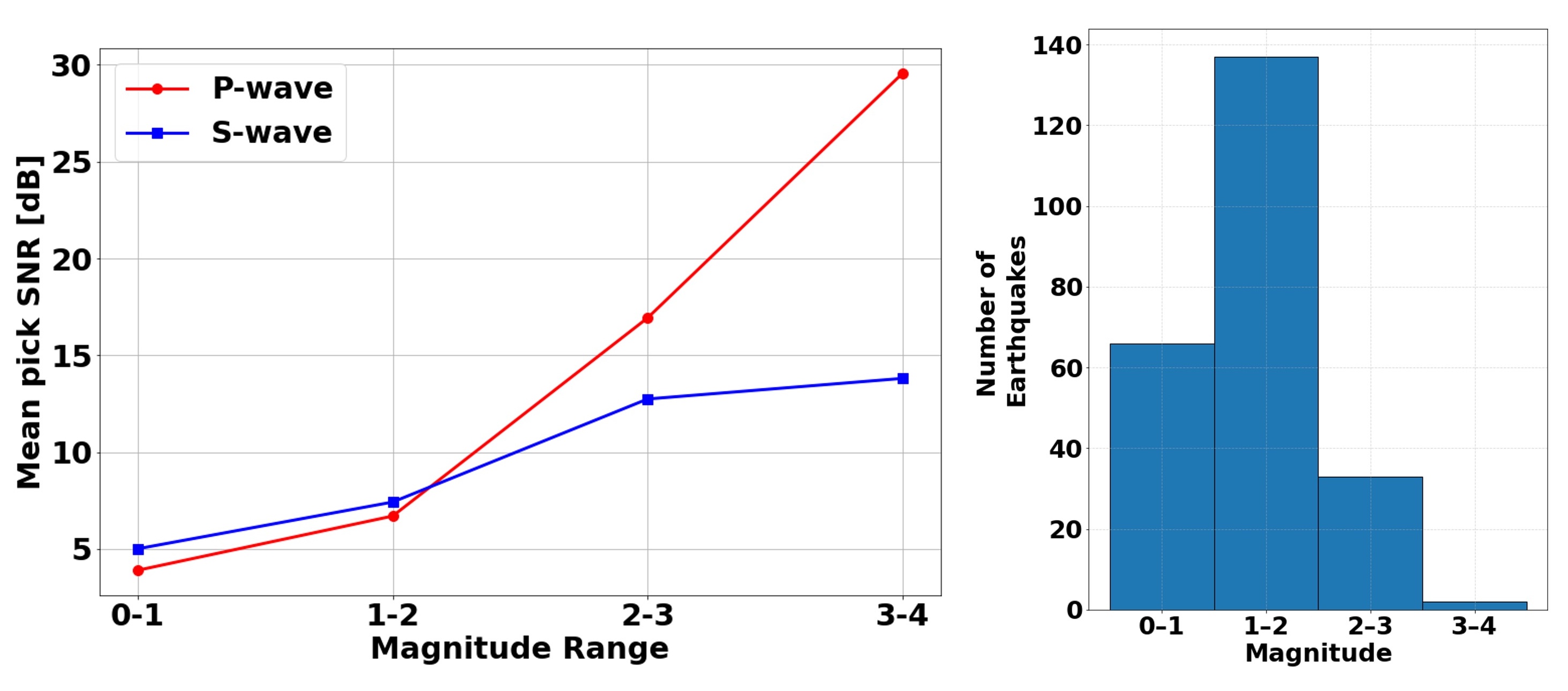}
\caption{Signal-to-noise ratio (SNR) level and magnitude distribution of picked local-regional earthquakes using PhaseNet-DAS on the Ridgecrest array for events recorded between May 27 and July 31, 2023.}
\label{fig:figure3new}
\end{figure}

\begin{figure}[ht]
\centering
\includegraphics[width=1.0\linewidth]{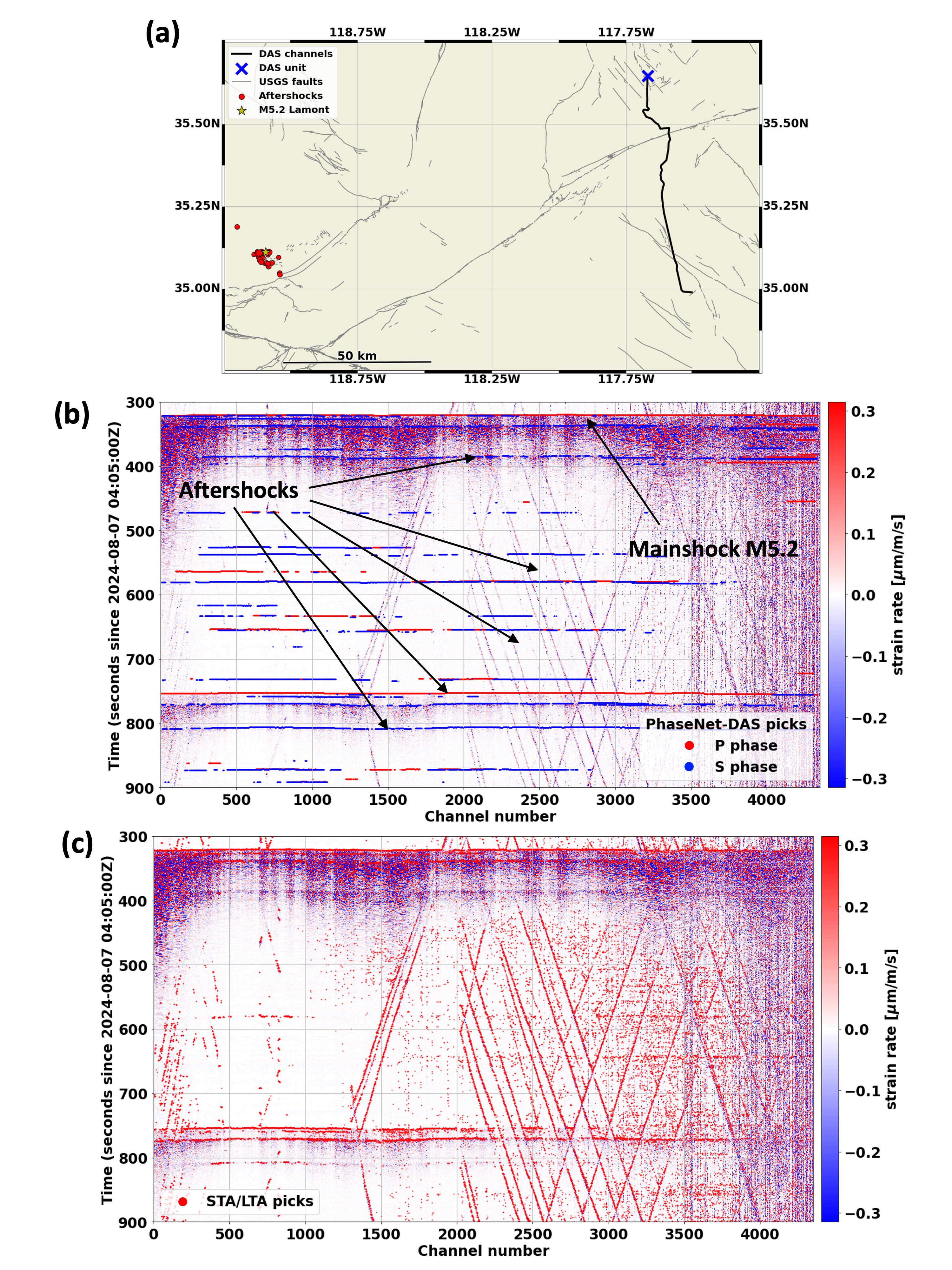}
\caption{Real-time picks from the M5.2 Lamont earthquake. (a) Map of the array and of the main shock location (yellow star) and of its aftershock (red dots). (b) DAS streamed data with the corresponding phase picks. (c) Traveltime picks obtained by a thresholding approach of STA/LTA curves.}
\label{fig:figure4}
\end{figure}

\subsection*{Supplementary materials}
Supplementary Figure S1 shows a Jiggle screen depicting a local M2.36 event (event ID 41153760) that occurred close to the Ridgecrest DAS array where P- and S-wave arrivals can be distinguished on six channels. Supplementary Figure S2 illustrates how lowering the STA/LTA detection threshold for DAS data increases the number of mispicks, primarily due to vehicle-induced deformations and instrument noise. Supplementary Movie S1 demonstrates the performance of PhaseNet-DAS on a rolling window for the M3.4 event shown in Figure~\ref{fig:figure3}a. Supplementary Movie S2 shows PhaseNet-DAS applied to a 70-second rolling window during the M5.2 Lamont sequence.

\end{document}